\documentclass[acmsmall]{acmart}

\renewcommand\footnotetextcopyrightpermission[1]{} 

\usepackage[utf8]{inputenc}

\usepackage{savesym}
\savesymbol{bigtimes}
\usepackage{mathabx}
\restoresymbol{MATHABX}{bigtimes}

\usepackage{natbib}
\setcitestyle{authoryear,open={(},close={)}}

\usepackage{xcolor}
\usepackage[]{algorithm}
\usepackage[noend]{algpseudocode}
\floatname{algorithm}{Listing}
\usepackage{xpatch} 
\makeatletter
\xpatchcmd{\algorithmic}
  {\ALG@tlm\z@}{\leftmargin\z@\ALG@tlm\z@}
  {}{}
\makeatother

\usepackage{moresize} 
\usepackage{listings}
\lstdefinestyle{static-analysis}{
  basicstyle=\scriptsize\ttfamily\color{black},
  moredelim=**[is][\color{red}]{@}{@},
}
\usepackage{caption}
\usepackage{subcaption}
\usepackage{booktabs}
\usepackage{array}
\usepackage{adjustbox}
\usepackage{multicol}
\usepackage{multirow}
\usepackage{pgf}
 \pgfmathsetseed{\number 5}
\usepackage{tikz}
\usetikzlibrary{arrows,positioning,shapes.geometric,calc,calendar,fpu,decorations,decorations.pathreplacing}
\usepackage{wasysym}

\newcommand{\code}[1]{\texttt{#1}}
\newcommand{\pipe}{\,|\,}

\newcommand{\FAST}[1]{\textsc{FAST}}

\title{%
    Language Support for Adaptation:
    Intent-Driven Programming in \FAST{}
}

\begin{abstract}
Historically, programming language semantics has focused on assigning
a precise mathematical meaning to programs.  That meaning is a function from the program's input domain to its output domain determined solely by its syntactic structure.  Such a semantics, fosters the development of portable applications which are oblivious to the performance characteristics and limitations (such as a maximum memory footprint) of particular hardware and software platforms.
This paper introduces the idea of \emph{intent-driven programming} where the meaning of a program additionally depends on an accompanying {\em intent} specification expressing how the ordinary program meaning is dynamically modified during execution to satisfy additional properties expressed by the intent. These include both  \emph{intensional} properties---e.g., resource usage---and \emph{extensional} properties--- e.g., \emph{accuracy} of the computed answer.

To demonstrate the intent-driven programming model's value, this paper presents a general-purpose intent-driven programming language---called \FAST{}---implemented as an extension of Swift. 
\FAST{} consists of an intent compiler, a profiler, a general controller interface and a runtime module which supports interoperation with legacy C/C++ codes. 
Compared to existing frameworks for adaptive computing, \FAST{} supports dynamic adaptation to changes both in the operating environment and in the intent itself, and enables the mixing of procedural control and control based on feedback and optimization.
\end{abstract}

\begin{document}

\author[Yang]{Yao-Hsiang Yang}
\affiliation{Rice University}
\email{yao-hsiang.yang@rice.edu}

\author[Duracz]{Adam Duracz}
\affiliation{Rice University}
\email{adam.duracz@rice.edu}
\orcid{0000-0003-4175-4020}

\author[Bartha]{Ferenc A. Bartha}
\affiliation{Rice University}
\email{barfer@math.u-szeged.hu}

\author[Sai]{Ryuichi Sai}
\affiliation{Rice University}
\email{ryuichi@rice.edu}

\author[Pervaiz]{Ahsan Pervaiz}
\affiliation{University of Chicago}
\email{ahsanp@uchicago.edu}

\author[Barati]{Saeid Barati}
\affiliation{University of Chicago}
\email{saeid.barati157@gmail.com}

\author[Nguyen]{Dung Nguyen}
\affiliation{Rice University}
\email{dxnguyen@rice.edu}

\author[Cartwright]{Robert Cartwright}
\affiliation{Rice University}
\email{cork@rice.edu}

\author[Hoffmann]{Henry Hoffmann}
\affiliation{University of Chicago}
\email{hankhoffmann@cs.uchicago.edu}

\author[Palem]{Krishna V. Palem}
\affiliation{Rice University}
\email{palem@rice.edu}

\makeatletter
\let\@authorsaddresses\@empty
\makeatother

\maketitle
\thispagestyle{empty}

\section{Introduction}

Recent proposals allow developers to create self-adaptive applications~\cite{oreizy1999architecture,kephart2003vision} while abstracting platform-specific monitoring and resource allocation logic~\cite{sampson2011enerj,imes2016bard,adaptive2017taylor,mishra2018caloree}. These systems provide a portable approach to constructing adaptive applications, but they are rigid with respect to the kind of adaptation they support and the way that the desired behavior is specified. 
Much of the recent research targets adaptation to meet energy constraints. While approaches exist that support re-targeting adaptation to meet latency constraints~\cite{imes2016bard,adaptive2017taylor,mishra2018caloree}, these frameworks lack a general-purpose mechanism (a language) for communicating and directly manipulating---in the program---that adaptive capability. Existing language-level support for adaptation~\cite{salvaneschi2013analysis} focuses on different mechanisms for introducing variability in programs, and on composing adaptive components. 
To the best of our knowledge, \emph{no existing system makes both the goal and means of adaptation general, first-class members of a programming language.}

Therefore, we propose a novel \emph{intent-driven programming} model that extends traditional general-purpose programming languages with new syntax to express \emph{intents} and a library API for identifying, monitoring, and manipulating the program's adaptive aspects. 
An intent (1) expresses the \emph{goal} of adaptation as a constrained optimization problem, (2) identifies variables (\emph{knobs}) in the application code or platform configuration that can be modified to achieve the goal, and (3) identifies values (\emph{measures}) in the application and hardware platform as elements of the objective function and constraint. An intent-driven program executes on top of a runtime that uses the intent to compute a schedule of knob settings that meets the optimization constraint while providing optimal behavior with respect to the objective function. In contrast to existing systems, the intent is \emph{general} and \emph{dynamic}---for example, intents may comprise arbitrary arithmetic expressions over the measures---and may change dynamically. The space of configurations that is available for adaptation can be changed dynamically from within the program, supporting use cases that require mixing procedural control with the automatic adaptation provided by \FAST{} using feedback and optimization.  This paper contributes:
\begin{itemize}
    \item 
        A precise definition of the intent-driven programming model, which enables the portable development of adaptive applications, by letting users express intents alongside their application code, which guide the dynamic reconfiguration of the system.
    \item 
        A description of the \FAST{} architecture, an instance of the intent-driven programming model built on top of the Swift programming language. FAST generalizes several types of existing frameworks for building adaptive systems, by supporting flexible intents written in terms of both application measures and platform measures, requiring adaptation in both application configuration and platform configuration.
    \item
        An experimental evaluation of the \FAST{} system on three example applications, executed on an embedded system. The evaluation demonstrates the ability to adapt dynamically to changes in operating conditions and in intent, including dynamic, programmatic manipulation of the configuration space. Benefits of this capability in \FAST{} compared to two other tools (PowerDial and OpenTuner) from the autotuning domain are also highlighted in the evaluation.
\end{itemize}


\section{Background and Related Work}
\label{sec:related-work}

Although no existing system with first-class language support for expressing intents is general enough to express the iterative, dynamically adaptive applications for which \FAST{} is designed, elements of the \FAST{} architecture can be found across a number of different domains. This section reviews the main related works, and discusses how they relate to \FAST{}. Table~\ref{tab:comparison-autotuning} compares \FAST{} to several systems from the autotuning domain (Section~\ref{sec:autotuning}), where the closest related work can be found.

\newcolumntype{R}[2]{%
    >{\adjustbox{angle=#1,lap=\width-(#2)}\bgroup}%
    r%
    <{\egroup}%
}
\newcommand*\rot{\multicolumn{1}{R{45}{1em}}}
\begin{table}
    \centering
    \begin{tabular}{@{}lllllll@{}}
        \toprule
         & \rot{PetaBricks} & \rot{Green} & \rot{EnerCaml} & \rot{OpenTuner} & \rot{PowerDial} & \rot{FAST} \\
        \midrule
        User-defined intents & \CIRCLE{} & \LEFTcircle{} & \LEFTcircle{} & \CIRCLE{} & \CIRCLE{} & \CIRCLE \\
        Programming language integration & \CIRCLE{} & \Circle{} & \CIRCLE{} & \Circle{} & \Circle{} & \CIRCLE{} \\
        Dynamic change to intent & \Circle{} & \Circle{} & \Circle{} & \Circle{} & \LEFTcircle{}{} & \CIRCLE{} \\
        Dynamic change to configuration space & \Circle{} & \Circle{} & \Circle{} & \Circle{} & \Circle{} & \CIRCLE{} \\
        Non-uniform configuration space & \Circle{} & \Circle{} & \Circle{} & \CIRCLE{} & \Circle{} & \CIRCLE{} \\ 
        Integrated dynamic control & \Circle{} & \CIRCLE{} & \Circle{} & \Circle{} & \CIRCLE{} & \CIRCLE{} \\
        General controller and/or optimizer concept & \Circle{} & \Circle{} & \Circle{} & \CIRCLE{} & \CIRCLE{} & \CIRCLE{} \\
        Energy-aware & \Circle{} & \CIRCLE{} & \CIRCLE{} & \CIRCLE{} & \CIRCLE{} & \CIRCLE \\
        \bottomrule
    \end{tabular}
    \caption{Comparison between different autotuning systems.}
    \label{tab:comparison-autotuning}
\end{table}

\subsection{Approximate computing}

Approximate computing comprises various techniques to trade off the quality of a computation's results for some other aspect of the computation, such as performance, latency or energy efficiency. The generality of the \FAST{} architecture in part stems from delegating the definition of application-specific notions such as quality to the programmer. In a \FAST{} application, quality is just a measure, and requires no distinguished treatment by the runtime. Thus, \FAST{} is flexible enough to constitute a general framework for expressing and automatically choosing among these different design choices. For example, loop perforation \cite{loop-perforation} can be implemented in \FAST{} by making the loop stride an application knob. General surveys of this field were given by \citet{approximate2013han,approximate2016xu,survey2016mittal}. The PowerDial system constructs knobs at compile-time and dynamically adjusts their settings at runtime to maintain a user-specified goal~\cite{hoffmann2011dynamic}.  JouleGuard coordinates application-level adaptation (like that done with PowerDial) with system-level resource management to meet energy guarantees \cite{hoffmann2015jouleguard}.  While similar to \FAST{}, PowerDial and JouleGuard are much less flexible as they do not provide users with a way to manipulate the knobs once the program is running. In contrast, \FAST{} programmers can dynamically change goals (Section~\ref{sec:intent-specifications}) and available knobs (Sections~\ref{sec:library-knob-type} and ~\ref{sec:x264-use-case-explicit-higher-quality-demand}) without stopping and recompiling.

\subsection{Autotuning Systems and Languages}
\label{sec:autotuning}

PetaBricks \cite{tuning2009ansel} allows users to provide different algorithmic choices, from which an autotuner finds an optimal combination. Its main purpose is to optimize the time efficiency of code automatically. Compared to \FAST{}, it does not have a general notion of a measure and makes choices at compile time.  Unlike PetaBricks, \FAST{} automatically adapts to dynamic events while allowing programmers to programmatically change intent and available configuration space.

Green \cite{tuning2010baek} does support programs that recalibrate themselves during execution. It provides two kinds of constructs: loop approximation and function approximation. Green also provides a statistical guarantee for the resulting quality of service. However, Green assumes a monotonic relationship between quality degradation and energy efficiency, and this information may not be readily available to a user. Unlike \FAST{}, Green does not have a general notion of knob and its controller is hard-wired. 

EnerCaml \cite{tuning2014ringeburg} extends OCaml with an approximation annotation system, a profiling system and an autotuner. The annotation system allows users to provide approximate versions for every expression. The profiling system calculates a single user-specified quality of a function's result (which corresponds to a \FAST{} measure). In contrast to the general measure monitoring facilities of \FAST{}, EnerCaml estimates energy savings by calculating the ratio of approximated operations to approximable ones. The autotuner calculates the Pareto frontier to see the trade-off between the estimated energy consumption and the quality of the result. EnerCaml is the first language designed to explore the trade-off between different approximation strategies. However, its concept of autotuning is restrictive, focusing only on energy savings, and does not have a general notion of measure. Further, it is designed for exploring different configurations at the prototyping stage and, unlike \FAST{}, does not allow the system to adapt to environmental changes that occur during execution.

To the best of our knowledge, OpenTuner \cite{tuning2014ansel} is the only currently available autotuning framework with a general notion of measure and search technique. It lets the user define both constrained and unconstrained objectives. OpenTuner allows users to explore the online behaviour of previously unexplored configurations. However, OpenTuner does not take into account the possible error between run time and compile time performance. Therefore, the control of iterative (e.g. streaming) applications cannot be accommodated with OpenTuner in a straightforward manner. 

\subsection{Hyperparameter Optimization}

Hyperparameter optimization is a two-level framework that tunes hyperparameters of a learning algorithm, to achieve better performance. Many standard optimization heuristics, such as simulated annealing and genetic algorithms, could be employed for this purpose. Currently, Optunity \cite{hyper2014claesen} is the only general library which supports hyperparameter tuning. \FAST{} similarly includes a unconstrained global optimization module which can be straightforwardly extended to support common hyperparameter optimization heuristics. For constrained optimization, Capri \cite{capri2017biswas} offers a general approach to find optimal hyperparameters of a program with input features taken into account.


\section{Intent-Driven Programming}
\label{sec:intent-driven-programming}

Adaptive software changes its behavior to meet a goal (intent) despite changes in external operating conditions. To do so, the software must detect that the current behavior deviates from the intent, determine the action necessary to recover that behavior, and implement that action. In the intent-driven programming model, these capabilities are enabled by:

\begin{itemize}
    \item
        a \textbf{library API} that lets the system control the application state (application knobs), with feedback (measures) to guide the controller, and delineate the scope where measures should be recorded (optimize);
    \item
        an \textbf{intent specification} that declares the knobs and measures that should be considered by the system, defines permissible assignments (ranges) for these knobs, and defines the goal of adaptation, in the form of a constrained optimization problem: to minimize or maximize an objective function, subject to a constraint;
    \item 
        a \textbf{runtime} that monitors the state of the application and platform as seen through the measures, computes the schedule of configurations needed to achieve the intent, and reconfigures the system accordingly, by changing the values of knobs.
\end{itemize}

We will see a concrete instance of the programming model in Section~\ref{sec:fast-architecture}, with details for each of the above components. The following subsections describe the three main concepts of the programming model, which connect these components. Table~\ref{tab:concepts} summarizes their relationships.
\begin{table}[tbp]
    \centering
    \begin{tabular}{
        @{}>{\raggedright}p{8em}>{\raggedright}p{9em}>{\raggedright}p{9em}p{9em}@{}
    }
        \toprule
            & Knob
            & Measure
            & Intent \\
        \midrule
        Library API 
            & Exposes its value to the application
            & Exposes its value to the runtime
            & Defines its~~ scope \\
        \addlinespace
        Intent specification
            & Defines its range
            & Declares it
            & Defines it \\
        \addlinespace
        Runtime	
            & Controls its value
            & Observes its value
            & Attempts to satisfy it \\
        \bottomrule
    \end{tabular}
    \caption{Main concepts of the programming model.}
    \label{tab:concepts}
\end{table}

\subsection{Knobs}
\label{sec:knobs}

A knob is a piece of program state that the runtime can modify to meet the user-defined intent. Knobs are classified as either \emph{platform knobs}, such as the CPU clock frequency and the number of cores on which the application's threads may be scheduled, or \emph{application knobs} such as the number of times to run an iterative algorithm, the minimum precision of an approximation scheme, or an identifier that selects among a set of possible algorithms to solve a particular sub-problem in the application. 

To be useful in an intent-driven program, the values that a knob can be assigned to (its range) should represent different \emph{trade-offs} between measures which are relevant to the intent. For example, increasing the platform knob that controls the CPU clock frequency can gain a lower processing latency at the expense of a higher energy consumption. Similarly, decreasing an application knob that controls the threshold of an approximation scheme can gain a higher output precision at the expense of a higher processing latency and energy consumption. An intent-driven program with knobs that trade precision for some other measure can be seen as an instance of the approximate computing paradigm~\cite{palem2014inexactness,approximate2016xu,schlachter2018trading}. Early foundational work on this topic can be fond in~\cite{palem2003computational,palem2005energy,chakrapani2008probabilistic}.

Though the two classes of knobs (platform vs. application) may achieve their trade-offs very differently, they are treated identically by the runtime. They allow it to reconfigure the system so that the measures change in such a way that the constraint (provided in the intent specification) is achieved.

We will refer to a defined set of knob bindings as a \emph{configuration}.

\subsection{Measures}
\label{sec:measures}

A measure is a part of the application or system state that is relevant to the intent. Examples of application measures include the output bit-rate of a compression algorithm, or the error of an approximation scheme. Examples of platform measures include input processing latency and energy consumption.
By exposing such state as a measure, it becomes available to the runtime as feedback. Based on this feedback, and on the intent declaration, the runtime can compute the knob settings necessary to meet the intent.

\subsection{Intents}
\label{sec:intents}

An intent is an encoding of desired program behavior, expressed in terms of measures. Generalizing existing work on adapting to meet latency constraints~\cite{imes2015poet}, \FAST{} intents take the form of a constrained optimization problem:
\begin{equation}\label{eqn:constrained-optimization-problem}
    \textup{opt}(f(m_0,\ldots,m_n)) \textup{ such that } 
    \left\{
        \begin{array}{l}
        m_c = g \\
        \forall k \in K.~ v_k \in D_k
        \end{array}
    \right.
\end{equation}
\noindent where $\textup{opt} \in \{ \min , \max \}$, $f$ is the objective function, $M = \{m_0, \ldots, m_n\}$ is the set of measures, $m_c \in M$ is the constraint measure, $g$ is the constant \emph{goal}, $K$ is the set of knobs, $v_k$ the value of knob $k$ and $D_k$ the domain of $k$. 
Solutions to the optimization problem can be expressed in terms of knobs, since the values of measures implicitly depend on the configuration of the system.
An execution of a system will be said to \emph{meet} the intent when the constraint measure $m_c$ is close to the goal $g$, while optimizing the objective function $f$ according to the optimization type. Figure~\ref{fig:comparison-fast-oracles} shows examples of such an execution. In Figure~\ref{fig:comparison-fast-oracles}a, the constraint measure \code{performance} and goal are illustrated by orange and green lines, respectively. In Figure~\ref{fig:comparison-fast-oracles}b, the objective function \code{quality} is illustrated by an orange line.

It should be noted that, from the perspective of the programming model, alternative formulations of intents are possible. 
For example, an intent could be a specification of a safety property that the runtime must seek to preserve with some probability.
The investigation of such alternatives is part of our future work.


\section{\FAST{} Architecture}
\label{sec:fast-architecture}

The \FAST{} system architecture is an instance of the general programming model, implemented as an extension of Apple's Swift language~\cite{swift}. This choice makes \FAST{} statically typed and memory-safe with predictable performance thanks to the automatic storage management (reference counting) and evaluation strategy (strict). As such, it is suitable for implementing soft real-time applications, making implicit system behavior such as latency or energy consumption easier to control reliably. However, the intent-driven programming model is by no means restricted to the imperative, object-oriented paradigm, since any language in which the rebinding of variables during execution is meaningful can serve as a basis for intent-driven programming.

\begin{figure}[t]
    \centering
    \hspace*{-2mm}\includegraphics[width=1.05\linewidth]{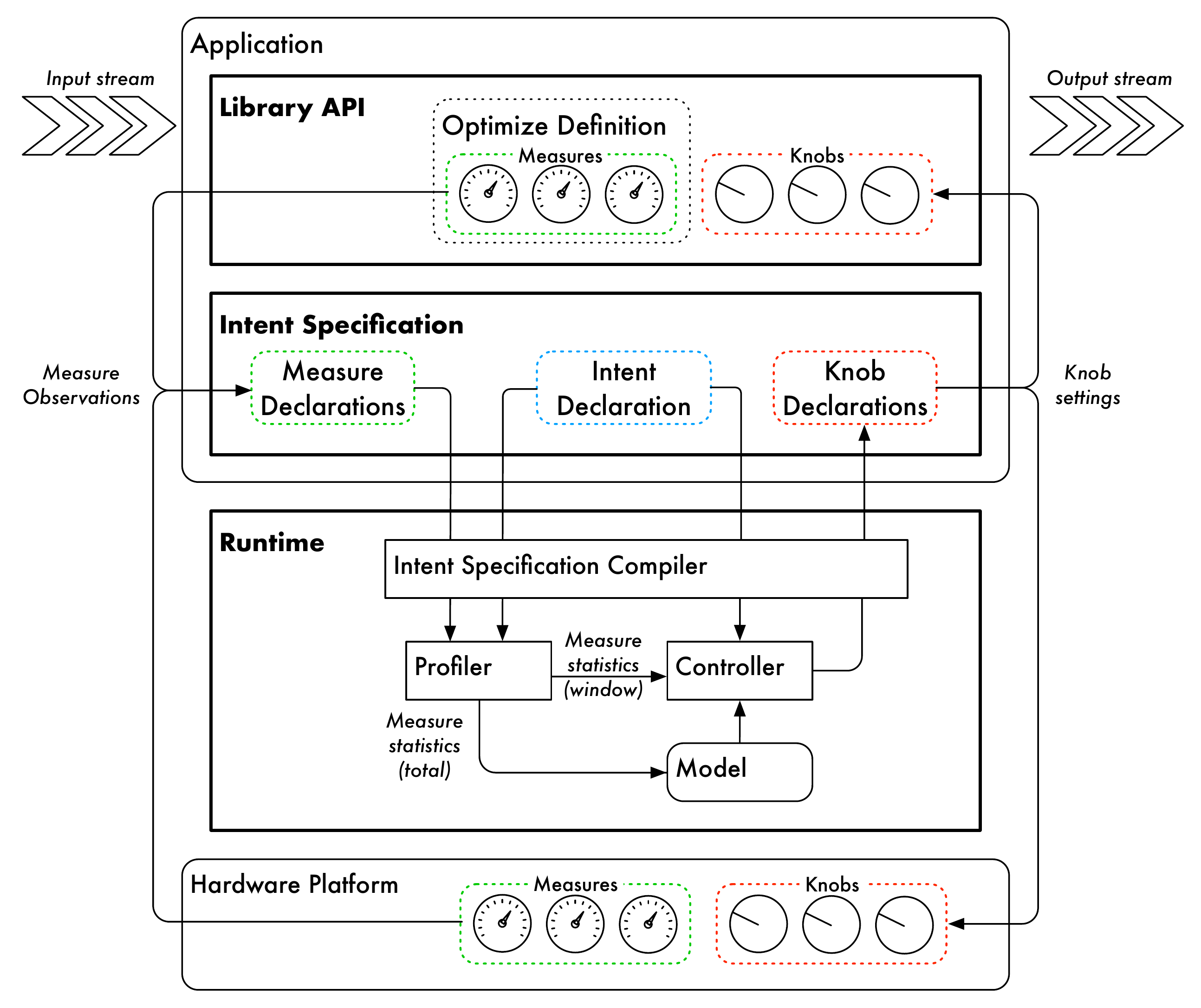}
    \caption{The FAST Architecture}
    \label{fig:fast-architecture}
\end{figure}

Figure~\ref{fig:swift-code-incrementer} shows a small Swift program that endlessly increments a variable \code{x} by a certain \code{step}, up to a certain \code{threshold}. We will use this program to illustrate \FAST{}'s user interface, which consists of a library API that enables \FAST{}'s runtime to control an application, and a domain-specific language for specifying the intent.

\begin{figure*}

\centering

    \begin{subfigure}[t]{0.25\textwidth}
        \begin{lstlisting}
        
let step = 1
let threshold = 2000000
while(true)
{
  var x = 0.0
  
  while(x < threshold)
  {
    x += step
  
  }

}
        \end{lstlisting}
        \caption{Swift code}
        \label{fig:swift-code-incrementer}
    \end{subfigure}%
    \begin{subfigure}[t]{0.42\textwidth}
        \begin{lstlisting}
import FAST
#let step =# Knob("step", #1#)
#let threshold =# Knob("threshold", #8000000#)
optimize#(#"incrementer", [threshold, step]#)# 
#{
  var x = 0.0#
  var operations = 0.0
  #while(x < threshold#.get()#)
  {
    x += step#.get()#
    operations += 1
  }
  #measure("operations", operations)#
}#
        \end{lstlisting}
        \caption{Swift (gray) and instrumentation (black)}
        \label{fig:swift-code-incrementer-instrumented}
    \end{subfigure}%
    \begin{subfigure}[t]{0.3\textwidth}
    \begin{lstlisting}
intent incrementer 
  min(energy*energy/operations) 
  such that latency == 0.1
measures
  latency: Double 
  operations: Double 
  energy: Double
knobs
  step = [1,2,3,4]
  threshold = 
    [2000000,5000000,8000000]
  coreFrequency = [300,1200]
  such that 
    threshold/step > 700000
    \end{lstlisting}
    \caption{Intent specification}
    \label{fig:intent-specification-incrementer}
    \end{subfigure}
    
\caption{The incrementer application}
\label{fig:incrementer}
\end{figure*}

Figure~\ref{fig:fast-architecture} illustrates the \FAST{} architecture, where three interacting components together constitute the \FAST{} runtime: an intent specification compiler, a profiler and a controller. 
The intent specification compiler (Section~\ref{sec:compiler}) translates the programmer's intent (Section~\ref{sec:intent-specifications})---specified through the library API (Section~\ref{sec:library-knob-type})---to an efficient representation, suitable for real-time processing.
The profiler (Section~\ref{sec:profiler}) computes measure statistics: total statistics offline to construct controller models; windowed statistics online as feedback to the controller  (Section~\ref{sec:controller}).

The next two sections describe the programmer interface, which consists of the library API (Section~\ref{sec:library-knob-type}), and the intent specification language (Section~\ref{sec:intent-specifications}).

\subsection{Library API}
\label{sec:library-api}

The library API consists of a type (\code{Knob}), and two functions (\code{measure} and \code{optimize}). The API is used to \emph{instrument} applications. This involves providing hooks into the application that the runtime can use to observe and adapt the application state, and identifying the portion of code that should be monitored by the runtime during execution.
Figure~\ref{fig:swift-code-incrementer-instrumented} shows the instrumented Swift code after a programmer has identified suitable values (measures), variables (knobs) and loop (optimize) in the base code, and wrapped them in the corresponding function or constructor calls.

\subsubsection{The Knob Type}
\label{sec:library-knob-type}

From the user's perspective, the type \code{Knob<T>} is a type-safe immutable cell that replaces a constant in the base program, and enables the runtime to adapt the cell's value. In our example, \code{threshold} and \code{step} become knobs and their original values are used to initialize the \code{Knob} type. 
These \emph{reference values} make it possible to compile and execute the instrumented application as a normal Swift program, with the original semantics. Thus, reference values represent the knobs' values if no runtime adaptation is possible. 
The \code{Knob} initializer also takes a string to identify the knob to the runtime.

Because knobs are mutated by the runtime between executions of the main processing loop (Section~\ref{sec:optimize}) they must be declared in an enclosing scope, meaning that, in a terminating application, they remain defined after the routine is executed for the last time.
However, the variable encapsulated by the \code{Knob} type remains referenced also when execution reaches the end of the scope where the knob was declared, since the runtime must have access to such a reference to perform its adaptation. Thus, to avoid memory leaks, the runtime must use only weak weak references to knobs.

The Knob type provides a novel mechanism to combine programmatic manipulation of the application configuration with the runtime's continuous control in the form of two methods: \code{restrict} and \code{control}.
For example, the programmer can enable or disable certain knobs, based on knowledge outside of the controller's domain. To this end, the Knob type's \code{restrict} method allows the developer to explicitly define a range of values for a particular \code{Knob}. Passed to \code{restrict} in the form of an array, the runtime uses this range to constrain configuration space available to the controller for adaptation.
Calling the method without any arguments fixes the \code{Knob} to the value it had at the time of the method call.
The Knob type's \code{control} method can be used to remove any restrictions from previous calls to \code{restrict}, making available to the controller the complete set of values for a \code{Knob} specified by the \code{knobs} section of the active intent.

\subsubsection{The Measure Function}
\label{sec:library-measure-function}

The \code{measure} function provides a view of the application's state as feedback to the controller. In our example, a new variable \code{operations} is added to the application to inform the runtime about the number of times that \code{x} has been incremented.
The \code{name} that is passed to the \code{measure} function is used by the runtime to correlate the given \code{value} with the corresponding measure in the intent specification.

\subsubsection{The Optimize Function}
\label{sec:optimize}

The \code{optimize} function replaces the outer \code{while} loop in our example. It takes as parameters: a name (\code{incrementer}), a list of knobs that the runtime should use for control, and a routine (block of code) that should be monitored to provide feedback to the controller, and an optional \emph{window size} over which measure statistics are computed.

The runtime obtains the feedback from the routine through side-effecting function calls, and thus the type of the routine is \code{Void -> Void}.
The window size segments the sequence of iterations into conceptual computation blocks. For a sufficiently small window size, the application behavior is assumed to be sufficiently uniform that average measure values of the previous window are representative of the iterations of the next window. On the other hand, the window size should be large enough that the windowed statistics filter out transients, caused by rare and unsystematic events.
Thus, the window size is typically identified by the application developer, who will be a domain expert, familiar with the expected behavior of the application under different conditions.

\subsection{Intent Specification}
\label{sec:intent-specifications}

\begin{figure}
    \centering
    \[\begin{array}{@{}c@{\hspace{2mm}}c@{\hspace{2mm}}l@{\hspace{2.5mm}}c@{\hspace{2.5mm}}l@{}}
        s & \in & \textsf{Spec} & ::= &
            \code{intent } n \code{ } o \code{(} e \code{)} 
            \code{ } [ \code{ such that } m \code{ == } c \code{ } ] \\
          &     &               &     & 
            \code{measures } 
            \{ \code{ } 
            m_i : t_i 
            \code{ } \}^{1 \leq i \leq |M_s|} \\
          &     &               &     & 
            \code{knobs } 
            \{\code{ }
            k_i \code{ = } e_i 
            \code{ } [ \code{ reference } c_i \code{ } ]
            \code{ } \}^{1 \leq i \leq |K_s|}
            \code{ } [ \code{ such that } e \code{ } ]\\
        o & \in & \textsf{Opt} & ::= &
            \code{min} \pipe \code{max} \\
        e & \in & \textsf{Expr} & ::= &
            c \pipe m \pipe f \langle e_i \rangle^{1 \leq i \leq |f|} \\
        c & \in & \textsf{C} & ::= & 
            \textit{constants} \\
        n & \in & \textsf{N} & ::= & 
            \textit{intent names} \\
        t & \in & \textsf{T} & ::= & 
            \textit{type names} \\
        m & \in & \textsf{M} & ::= &
            \textit{measure names} \\
        k & \in & \textsf{K} & ::= &
            \textit{knob names} \\
        f & \in & \textsf{F} & ::= &
            \textit{function names}
    \end{array}\]
    \caption{The \FAST{} Intent Specification Language}
    \label{fig:fast-isl-grammar}
\end{figure}

Figure~\ref{fig:fast-isl-grammar} shows a grammar for the \FAST{} Intent Specification Language. In this figure, $\{\cdot\}$ and $\langle\cdot\rangle$ denote a set and a sequence of elements, respectively. The arity of a function $f$ is denoted by $|f|$. The set of names of knobs declared in an intent specification $s \in \textsf{Spec}$ is denoted by $K_s$. The set of names of measures declared in $s$ is denoted by $M_s$.
Figure~\ref{fig:intent-specification-incrementer} is an example intent specification for the incrementer application in Figure~\ref{fig:swift-code-incrementer-instrumented}. An intent specification consists of two main parts: an encoding of an optimization problem, and a description of the degrees of freedom along which the system may operate. 

\subsubsection{Intent}

The \code{intent} section encodes an (optionally) constrained optimization problem and consists of five parts:
\begin{itemize}
    \item 
        The name of the \code{optimize} routine (\code{incrementer} in our example) that the intent should affect. This is correlated with the name passed to the \code{optimize} function.
    \item 
        The \emph{optimization type}, one of \code{min} or \code{max}.
    \item
        The \emph{objective function}, an expression in terms of the declarations of the \code{measures} section.
    \item
        The \emph{constraint measure}, the one from the \code{measures} section (\code{latency} in our example) that the runtime should control.
    \item
        The \emph{constraint goal}, the value (\code{0.1} seconds per iteration in our example) of the constraint measure that the runtime should achieve.
\end{itemize}

The intent section expresses a high-level specification of what it means for the application to perform well, in terms of the measures which, notably, can be both intensional and extensional.

\subsubsection{Measures}

The \code{measures} section declares measures that should be observed by the runtime. These declarations serve as an environment for the \code{intent} section. Currently, measures may only have the \code{Double} type, but any totally ordered type that supports the operations used in the objective function could be supported, and the constraint measure could be of any type for which equality is defined.

Measures correspond to observable platform signals and expressions in the base application, in terms of which the intent is expressed. Throughout program execution, their current value serves as feedback to the runtime.

\subsubsection{Knobs}

The \code{knobs} section defines the possible configurations, or \emph{configuration space} that the \FAST{} runtime can select from during execution to solve the optimization problem specified in the \code{intent} section. The \code{knobs} section does this through a set of knob definitions that each consist of a name, a \emph{range} expression that evaluates to a list of constants, and a reference value. The name associates the knob definition with \code{Knob} instance in the application. The range is a list of values that the knob can be set to, which must be of the same type as the knob instance's type parameter. The optional reference value is used to override the reference value passed to the \code{Knob} constructor (Section~\ref{sec:library-knob-type}) to provide an initial value for the knob when the application is executed without control, and in the very beginning of a controlled execution.

In addition to the basic knob ranges, the \code{knobs} section supports an optional \emph{knob constraint}---an arbitrary Boolean expression over the knobs---which makes it possible to specify non-uniform configuration spaces. The knob ranges together generate the cross product of the knob settings, that is, all possible configurations that the system supports. The knob constraint is then used to remove all configurations that do not satisfy it.
This is useful when there is some kind of dependency between the knobs. For example, there may be cases when certain combinations of knob settings are not meaningful, such as when a knob is a parameter for a sub-algorithm that is enabled or disabled by another knob, or when some global restriction affects what values a set of knobs can be set to simultaneously.
Another use for knob constraints is to prune out redundant configurations that expose analogous tradeoffs between the measures that the intent depends on. Such pruning can both save time and energy in profiling the system before it is deployed, and in computing schedules during runtime.

Knobs typically correspond to platform settings or constants in the base application, whose value determines some aspect of system behavior. In non-adaptive applications, such constants are judiciously chosen by the application developer to achieve reasonable expected-case behavior, or exposed to users as parameters. Improperly setting these constants is a notorious source of performance issues \cite{huang2015understanding,rabkin2013hadoop,Autoconf}.
By exposing knobs to the runtime as variables, the system allows its behavior (as observed through the measures) to adapt to changes in the intent, or in the operating environment and helps eliminate performance bugs due to poor choices of constants.

\subsection{Runtime}
\label{sec:architecture-runtime}

The main components of the runtime are: an intent specification compiler that makes the programmer's intent available to the rest of the system, a profiler that observes the behavior of the executing application, and a controller that uses these observations to configure the system so that the intent is met. 

\subsubsection{Intent Specification Compiler}
\label{sec:compiler}

Users specify intents in files separate from the Swift application code. Compilation is also separate: Swift is a compiled language, and thus the program is compiled into an executable before runtime; \FAST{} intent specifications are interpreted at runtime. This approach has some benefits, compared to expressing intents within the Swift source code: intent specifications can easily be passed to an external system, such as a profiler, or updated over a network, without the need for expensive recompilation or complex binary plugin architectures. The \FAST{} intent specification compiler translates intent specifications into a form that permits evaluation without significant interpretative overhead~\cite{carette2009finally}, by transforming abstract syntax into nested Swift closures.

\subsubsection{Profiler}
\label{sec:profiler}

The profiler collects statistics about measures. The statistics are used  to (1) construct models that the controller uses to predict configurations' measure values and (2) serve as feedback to the controller during execution (Section~\ref{sec:controller}). Figure~\ref{fig:model-incrementer} shows such a model for the incrementer application (Figure~\ref{fig:incrementer}). 

Statistics with two different horizons are collected. First, \emph{total} statistics are computed, based on all the observed values of each measure. These averages are representative of the over-all behavior of the system with respect to each measure, and are thus used to construct controller models. Second, \emph{window} statistics are computed, based on a sliding window of observed values. This window size, a parameter of the \code{optimize} function (Section~\ref{sec:optimize}), can be used to dampen noise in the measures, at some cost in the ability to react to legitimate abrupt changes in measure values. A small window size means that short spikes in measure values can cause oscillation in the controller, while a large window size can cause unacceptable lag in adaptation.

For each type of horizon, two statistics are computed: \emph{averages} are used in place of point samples of measure values; \emph{variances} are used for debugging purposes, to estimate the reliability of averages used in controller models. 
An important consideration when computing statistics in the context of iterative applications is that incremental (constant-time) algorithms~\cite{west1979updating} must be used, since the size of the set of observations over which statistics are computed is potentially unbounded.

\begin{figure}

\begin{subfigure}[t]{0.05\columnwidth}
    \centering
    \begin{tabular}{@{}r@{}}
        \toprule
        id \\
        \midrule
        0 \\
        1 \\
        2 \\
        3 \\
        4 \\
        5 \\
        6 \\
        7 \\
        \bottomrule
    \end{tabular}
\end{subfigure}%
\begin{subfigure}[t]{0.375\textwidth}
    \centering
    \begin{tabular}{@{}rrr@{}}
        \toprule
        step & threshold & coreFrequency \\
        \midrule
        1    & 10000     & 300	      \\
        4    & 10000     & 300	      \\
        1    & 20000     & 300	      \\
        4    & 20000     & 300	      \\
        1    & 10000     & 1200	  \\
        4    & 10000     & 1200	  \\
        1    & 20000     & 1200	  \\
        4    & 20000     & 1200	  \\
        \bottomrule
    \end{tabular}
    \caption{Knob Table}
    \label{fig:knob-table-incrementer}
\end{subfigure}%
\begin{subfigure}[t]{0.375\textwidth}
    \centering
    \begin{tabular}{@{}rrr@{}}
        \toprule
        energy & latency & operations \\
        \midrule
        6048055 &   0.017 & 10000 \\
        5367987 &   0.011 &  2537 \\
        10362040 &   0.031 & 19949 \\
        4311562 &   0.011 &  5025 \\
        3495722 &   0.008 & 10000 \\
        2587574 &   0.004 &  2537 \\
        4904005 &   0.012 & 19949 \\
        2729713 &   0.006 &  5025 \\
        \bottomrule
    \end{tabular}
    \caption{Measure Table}
    \label{fig:measure-table-incrementer}
\end{subfigure}

\caption{Controller model for the Incrementer application with eight configurations.}
\label{fig:model-incrementer}
\end{figure}

\subsubsection{Controller}
\label{sec:controller}

The \FAST{} runtime interacts with the controller component through a protocol with a single method that, given an intent specification and values for each declared measure, returns a \emph{schedule}.
A schedule is a function that, given a non-negative integer \emph{index}, returns a configuration.
A configuration is an object with an \code{apply} method, that can be used to reconfigure the system---that generally comprises both application and hardware.
This interface allows for many different types of controllers. The intent parameter determines the resource scheduling problem that the controller must solve, the measures enable feedback control, and the indexed form of the return type supports schedules that reconfigure the system at every iteration.

Next, we describe two example instance of this protocol.

\paragraph{Unconstrained Optimizing Controller}

This controller solves the optimization problem specified in the intent, without taking into account the additional constraint. By default it performs a grid search, and chooses the optimal profiled configuration. This applies to both streaming and non-streaming applications. This controller can be extended with meta-optimization heuristics~\cite{hyper2014claesen} to find optimal configurations at runtime when exhaustive search is not possible, such as when some knobs are continuous.

\paragraph{Constrained Optimizing Controller}
\label{sec:constrained-optimizing-controller}

\begin{figure}
    \centering
    \includegraphics[scale=1.65]{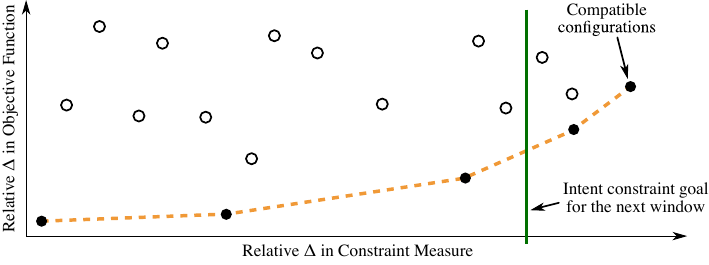}
    \caption{Constrained Optimizing Controller. A simplistic set of configurations is depicted in this figure. Configurations that appear in optimal schedules (in the case of minimizing the objective function) are on the lower convex hull in this coordinate system.}
    \label{fig:constrained-optimizer}
\end{figure}

This controller solves the optimization problem specified in the intent \emph{on average} across a window of iterations.
For \FAST{} intent specifications, this can be achieved robustly and efficiently using feedback control and optimization~\cite{Hellerstein2004,ControlSurvey}, based on a \emph{controller model} that consists of pre--recorded measure values (Figure~\ref{fig:measure-table-incrementer}) for each configuration that the system can be set to (Figure~\ref{fig:knob-table-incrementer}), which is obtained through profiling (Section~\ref{sec:profiler}). The optimal schedule is found by first identifying all constraint-compatible schedules, that is, collections of configurations that can be interleaved to meet the intent on average across a window, as illustrated in Figure~\ref{fig:constrained-optimizer}. Among these, the schedule that optimizes the objective function is selected for execution over the next window.


\section{Analysis of Intent-Driven Programs}
\label{sec:correctness}

Analyzing adaptive software is strictly harder than analyzing non-adaptive software. The difficulty arises due to both the number of dimensions along which the system can adapt and that adaptation happens over time.
Thus, extending a programming language with intents drastically alters its semantics, and also introduces several classes of bugs. Though a full treatment of testing and verification for intent-driven programs is too broad for one paper, the following sections briefly introduce a notion of cost semantics for intent-driven programming, define some simple static analyses that are useful in this context, and describe a testing approach for iterative intent-driven programs.


\subsection{Cost Semantics}

While many different cost semantics have been proposed, they share the property that the cost of an expression is compositional~\cite{blelloch1995cost, spoonhower2008cost}. In other words, an expression's cost is solely determined by the cost of each of its sub-expressions, and the rule that binds them together. If this algebraic system captures the real cost of computation across different machines with respect to a set of parameters, then we obtain a useful abstraction of real computational cost. Most existing cost semantics only model a specific kind of cost; e.g. time or memory. To extend such a semantics to an intent-driven language, we must further abstract away the the details of the computation's resource usage. We call this an \emph{asymptotic cost semantics}. As long as the resource is: (a) not reusable, (b) measurable for every single execution, and (c) its cost is additive, then the cost of a sequence of executions of a single expression can be modeled by a sequence of \emph{i.i.d.} random variables, conditional on the value of the expression's input. Further, the average of this sequence converges to some fixed value with an \emph{i.i.d} sequence of inputs, by the law of large numbers. We call this value the \emph{asymptotic cost} of an expression $e$ with respect to the underlying machine $M$, the resource $r$ and the distribution of the inputs $d$ and denote it by $C[e;M,r,d]$.

This semantics tries to capture the average (and thus cumulative) cost instead of the cost of each call. Intuitively, this means that the impact of compiler optimizations and external disturbances may be smoothed out in the long run, and the asymptotic cost semantics offers a more robust and manageable way to model the behavior for further tuning, in contrast to call-wise cost semantics and even relational cost semantics~\cite{cicek2017relational}.

A convenient property of an asymptotic cost semantics is that if we ``unroll'' the loop (i.e. this infinite sequence) $n$ times (denoting $e;e;\ldots;e$ by $e^n$), then:
\[
C[e^n;M,r,d^n] \leq nC[e;M,r,d],
\]
where the inequality would become strict only after some sort of (program) optimization. 
From this, we can further define a lower bound for our asymptotic cost:
\[
C_{L}[e;M,r,d] = \liminf_{n\rightarrow\infty} \frac{C[e^n;M,r,d^n]}{n} \leq C[e;M,r,d],
\]
which characterizes (gives an upper bound for) the best cost that (program) optimization could achieve asymptotically.

This notion of asymptotic cost semantics provides a natural way to understand \FAST{} program behavior.  Further, it provides a distinct advantage over other (non-asymptotic) cost semantics, since constrained optimal parameter tuning is easily achievable with it and almost always (that is, with probability one for all constrained non-degenerate cases which we will elaborate below) better than tuning each execution independently.
Let us consider the knob control problem in \FAST{}. Suppose we have an expression $e_k$, which depends on a parameter $k$ that corresponds to application knobs and a machine $M_{k'}$, which depends on a parameter $k'$ that corresponds to platform configuration knobs both defined previously in Section~\ref{sec:knobs}. Our cost functions $C[e_k;M_{k'},\cdot,d]$ represent measures defined in Section~\ref{sec:measures}. To minimize the asymptotic cost, we solve an infinite system corresponding to our intent defined in Section~\ref{sec:intents}: 

\begin{equation*}\label{infinite-opt}
\begin{aligned}
& \underset{\mathbf{k}, \mathbf{k'}}{\min}
& & \lim_{N\rightarrow\infty}\sum_{i=1}^{N}{\frac{1}{N}C[e_{k_i};M_{k'_i},o,d]} \\
& \textrm{subject to} 
& & \lim_{N\rightarrow\infty}\sum_{i=1}^{N}{\frac{1}{N}C[e_{k_i};M_{k'_i},r_l,d]} \preceq \mathbf{R}_l, \forall l, \\
\end{aligned}
\end{equation*}
where $o$ is the resource corresponding to the objective function in the intent, $r_l$ is the resource we want to constrain, and $\mathbf{R}_l$ is the constraint value. Notice here that, although the cost is defined differently for every different intent, the \code{restrict} API (Section~\ref{sec:library-knob-type}) merely changes the domain of optimization and thus (1) the original asymptotic cost still provides an upper bound after restriction and (2) all the information contained in the original cost function can be reused.

\noindent The above is equivalent to the following system:
\begin{equation*}\label{finite-approx-opt}
\begin{aligned}
& \underset{\mathbf{w}}{\min}
& & \sum_{k \in K}{C[e_k;M_{k'},o,d]w_{k,k'}} \\
& \textrm{subject to} 
& & A \mathbf{w} \preceq \mathbf{R}, \\
& & & \sum_{(k,k') \in K} w_{k,k'} =  1, \\
& & & w_{k,k'} \succeq 0. \\
\end{aligned}
\end{equation*}
where $A_{l,k}=C[e_k;M_{k'},r_l,d]$, $w_{k,k'}$ is the weight corresponding to each configuration $(k,k')$, $\mathbf{R}$ is the constraint value vector, and we can denote the corresponding optimal schedule $\{(k^*_i,k'^*_i)\}^{\infty}_{i=0}$. 
A uni-constraint version of this equivalence has been established and studied by \citet{imes2015poet}.

It is easy to see that:
\[\arraycolsep=1.4pt\def\arraystretch{1.75}
\begin{array}{cl}
     & \lim_{N\rightarrow\infty}\sum_{i=1}^{N}{\frac{1}{N}C[e_{k^*_i};M_{k'^*_i},o,d]} \\
\leq & \lim_{N\rightarrow\infty}\sum_{i=1}^{N}{\frac{1}{N}C[e_{k^{**}};M_{k'^{**}},o,d]} \\
=    & C[e_{k^{**}};M,o,d],
\end{array}\]
where $(k^{**},k'^{**})$ is a solution of the corresponding system without iterating:
\begin{equation*}
\begin{aligned}
& \underset{k}{\min}
& & C[e_{k};M_{k'},o,d] \\
& \textrm{subject to} 
& & C[e_{k};M_{k'},r_l,d] \preceq \mathbf{R}_l, \forall l. \\
\end{aligned}
\end{equation*}
The equality happens either when the system is unconstrained, or the system is degenerate. This inequality shows what we stated earlier: in the long run, optimal tuning with our asymptotic cost model almost always outperforms tuning with a non-asymptotic cost semantics (i.e., tuning executions independently).

\subsection{Static Analysis}
\label{sec:static-analysis}

The \FAST{} architecture separates the specification of programs from the specification of intents. The resulting flexibility comes at the cost of possible program errors, when the definitions in the two parts of a \FAST{} program are inconsistent. For example, for a program to be correct, uses of the Knob type (part of the library API described in Section~\ref{sec:library-knob-type}) must correspond to entries in the \code{knobs} section of an associated intent specification (described in Section~\ref{sec:intent-specifications}).
In addition, the \code{optimize} construct (Section~\ref{sec:optimize}) takes a list of knobs and passes them to the runtime. Using static analysis techniques, the system can provide meaningful feedback early during \FAST{} application development, and eliminate the possibility of certain types of runtime errors. This subsection presents three static analyses, illustrated in Figure~\ref{fig:static-analysis}, which are specific to the implicit programming model.

\subsubsection{Finding Unused Knobs}

Knobs defined in the intent specification but not declared as a Knob type are ignored by the runtime. On the other hand, declared knobs without configurations in the intent cannot be used in trade-offs while tuning the system.
This can be detected statically by collecting the list of knobs defined in the intent, and those declared in the \FAST{} application code, respectively. The analysis reports the difference, if any, between the two lists of knobs, and marks them as unused to aid in further analysis.

Define $K_{I}$ to be the set of knobs defined in the intent file, $K_{D}$ to be the set of knobs declared in the \FAST{} application code, $K_{O}$ to be the set of knobs passed to the optimize construct, and $K_{A}$ to be the set of knobs affecting the body of optimize construct. A knob $k \in K_{A}$ if there is a branch of execution which depends on the value of $k$.
Then the unused knobs $K_{UU}$ are defined as $(K_{D} \setminus K_{I}) \cup (K_{I} \setminus K_{D})$. The uncaptured knobs $K_{UC}$ are defined as $K_{D}-K_{O}$. The unaffected knobs $K_{UA}$ are defined as $K_{O}-K_{A}$.

\begin{figure}

\begin{subfigure}[t]{\textwidth}

    \hfill
    \begin{subfigure}[b]{0.5\textwidth}
        \begin{lstlisting}[style=static-analysis]
import FAST

let @uncaptured@ = Knob("uncaptured", 1)
let @unaffected@ = Knob("unaffected", 1)
let affected = Knob("affected", 1)

optimize("app", [@unaffected@, affected]) {
  var x = read(...)
  if (x < affected.get()) 
  {  sleep(@unaffected@.get())  }
  else 
  {  for i in 1..<10 { sleep(20) }  }
}
        \end{lstlisting}
        \caption{Application Code}
        \label{fig:application-code}
    \end{subfigure}
\begin{subfigure}[b]{0.45\textwidth}
    \begin{lstlisting}[style=static-analysis]
intent app 
  min(energy) 
  such that 
  latency == 0.1
measures 
  latency: Double energy: Double
knobs
  @unused@     = [1,2,3,4] reference 1
  @uncaptured@ = [1,2,3,4] reference 1
  @unaffected@ = [1,2,3,4] reference 1
  affected   = [1,2,3,4] reference 1
  
    \end{lstlisting}
    \caption{Intent Specification}
    \label{fig:intent-specification}
\end{subfigure}
\end{subfigure}
\caption{An Example Showing Three Kinds of Problematic Knobs}
\label{fig:static-analysis}
\end{figure}

\subsubsection{Finding Uncaptured Knobs}

In the case where a knob is declared, but is not passed to the optimize construct, it is not controlled by the system.
We say that these knobs are \emph{uncaptured}.
An analysis finds such knobs and emits a warning.

\subsubsection{Finding Unaffected Knobs}

Even when a knob is captured, the user may forget to use it, or inadvertently misuse it in the program.
Any tuning done by \FAST{} of such a knob will have no effect on the system, meaning that the runtime will be unable to exploit any trade-offs exposed by this knob.
Profiling can expose the presence of such knobs, which will correspond to near-identical entries in the measure table (Section~\ref{sec:profiler}). However, such dynamic analysis can be prohibitively expensive when the configuration space is large. 
A more efficient alternative is to use static analysis to identify such situations.
The analysis begins by building a data flow graph starting from the \code{optimize} construct. For each node in the graph, it computes the knobs that affect it.
Given the annotated graph, it is possible to compute the list of all effective knobs for the \code{optimize} construct, and issue a warning when this list is missing some declared knob.\\

This definition of an unaffected knob amounts to an all-or-nothing identification problem, and the optimization problem to be solved by the controller may be ill-defined in the presence of such knobs. Therefore, we did not include sensitivity in our static analyses. Surely, some form of sensitivity (how the initial condition will affect the solution) similar to the ``condition number'' of a linear system can be defined for our control problem, to detect whether the control system is functioning well. However, this would be computationally expensive, input-dependent and platform-dependent. On the other hand, finding a branch of code that possibly will not be controlled by some knob at runtime is a cheap solution that is both input-independent and platform-independent. To reiterate the main difference: the statistical approach is to detect whether the system is ill-``conditioned'' while our current approach is to find whether the system is ill-defined.

\subsection{Testing}

Traditional testing, based on hand-crafted test cases, becomes unfeasibly labor-intensive for intent-driven programs. Random testing, and its statistical perspective on correctness, scale to adaptive applications. Random testing produces a distribution of test cases that elicit all possible (including both representative and worst-case) behavior. Computing an evaluation criterion, which produces a set of possible outcomes, over a set of test cases produces a distribution of outcomes. Correctness can thus be phrased as a hypothesis over this distribution, say, that a test case picked from the test case distribution (with the desired coverage) produces a negative outcome with a small probability.

In this section, we make this approach to evaluation concrete for intent-driven applications, by describing how to generate test cases based on an intent specification, and by defining an application-agnostic evaluation criterion. We use this criterion to show that three example adaptive applications behave correctly on a test corpus, derived from the intent specification.

\subsubsection{Evaluation Criteria}

Implementing an intent-driven program starts with implementing a traditional program, which is done according to standard software development practice. That typically involves the implementation of automated tests. These tests should validate the correct behavior of the program under both normal and exceptional operating conditions, that is, for a set of program inputs and combinations of program parameter values that sufficiently reflect the use cases of the application.
These tests do not validate the correct behavior of the instrumented, intent-driven program. Crucially, the ability of the application to react to changes in the operating environment or user intent is not validated.

We devise a largely application-independent evaluation criterion that captures bugs influencing the overall system's adaptability. This criterion is constructed by comparing the tested application's execution trace---as controlled by \FAST{}---to an \emph{oracle}---constructed from a set of traces corresponding to application executions in fixed configurations. 

\subsubsection{Oracles}
We consider two types of oracles: A and B.

Oracle A represents a minimal requirement: that the adaptive application should do better than a non-adaptive system \emph{in some respect}. For example, embedded real-time systems are provisioned to deliver performance on---possibly rare---worst case inputs, at the cost of resources such as energy. A well constructed adaptive system---i.e., with the right instrumentation and  \emph{intent}---should achieve \emph{sufficient} performance (given the intent), while using fewer resources than the system that always assumes the worst case. This type of oracle could be constructed by looking at the controller model, and configuring the system according to its performance measure, or by computing a set of execution traces, based on the intent, and selecting the one with the best overall performance. This access to global, posterior knowledge is why these executions are called oracles.

Oracle B models ideal adaptive behavior. Such an oracle can be constructed from a set of fixed-configuration executions in an iteration-wise fashion. At every iteration, the oracle's measure values are chosen from those of the fixed-configuration execution that best meet the intent. An oracle for testing the constrained optimizing controller is \emph{close enough} to satisfying the equality constraint while optimizing the objective function.

To be clear, these two notions of oracle are approximations. They are intended to capture salient aspects of the application's adaptive behavior, while remaining tractable and simpler than the implementation they are modeling. Consequently, for example, there will be situations where an adaptive \FAST{} execution may perform better than Oracle B (for example, when no configuration comes close to satisfying the constraint).

When the application that is being tested contains uses of the Knob type's \code{restrict} API (Section~\ref{sec:library-knob-type}), valid oracle definitions must take into account the corresponding dynamic change to the configuration space. Oracles may only use those configurations that remain available to the runtime that is controlling the tested application. In other words, in the presence of the \code{restrict} API, valid oracles are parameterized by the model that is available at each iteration.

\subsection{Verdict Expressions}

Based on an oracle, we can construct a \emph{verdict expression} that maps a test case to one of a set of possible outcomes, such as \{PASS, FAIL\}. The definition of these outcomes depends on the chosen oracle, since, as discussed in the previous section, oracles may represent different reference points for the system under test. To compare an adaptive execution against an oracle, a notion of error is needed, and this notion will vary depending on what the oracle is designed to optimize.

Listing~\ref{fig:verdict-expression-a} shows how a verdict expression can be constructed based on Oracle A, and the three auxiliary definitions given below in terms of the execution $X$, which can be either Oracle or \FAST{}:
\begin{itemize}
    \item
        The mean absolute percentage error (MAPE) $E(X)$ of the constraint measure value for $X$, compared to the constraint goal.
    \item
        The cumulative objective function $F(X)$ of $X$.
    \item
        The global constraint measure error threshold $T$ of the constraint measure versus the constraint goal.
\end{itemize}

\begin{algorithm}
    \small
    \begin{algorithmic}[]
      \If{$E(\textup{Oracle}) > T$}
          \Comment{Oracle does not meet the constraint}
          \If{$E(\textup{FAST}) \leq T$}
            \Comment{FAST meets the constraint}
            \State PASS
          \Else
            \Comment{Neither FAST nor Oracle A meet the constraint}
            \State INVALID
          \EndIf
      \Else
          \Comment{Oracle meets the constraint}
          \If {$E(\textup{FAST}) \leq T$}
            \Comment{FAST meets the constraint}
            \If {$F(\textup{FAST})$ more optimal than $F(\textup{Oracle})$}
              \State PASS
            \Else
                \Comment{Oracle is more optimal than FAST}
                \State FAIL
            \EndIf
          \Else
            \Comment{FAST does not meet the constraint}
            \State FAIL
          \EndIf
      \EndIf
    \end{algorithmic}
    \caption{Verdict Expression Based on Oracle A}
    \label{fig:verdict-expression-a}
\end{algorithm}

Listing~\ref{fig:verdict-expression-b} shows how a verdict expression can be constructed based on Oracle B, and the following additional auxiliary definitions, given in terms of the executions $X_1$ and $X_2$, which can be either Oracle or \FAST{}:
\begin{itemize}
    \item
        The objective function advantage $A(X_1,X_2)$ which is defined as 0 when $F(X_1) \leq F(X_2)$ and defined as 
        $\frac{ F(X_1) - F(X_2) }{ \max(|F(X_1)|,|F(X_2)|) }$ 
        otherwise.
    \item
        The global constraint measure error threshold $T_E$ of the constraint measure versus the constraint goal, and the global objective function threshold $T_F$.
\end{itemize}

\begin{algorithm}
    \small
    \begin{algorithmic}[]
      \If{$E(\textup{FAST}) > T_E$}
        \Comment{FAST does not meet the constraint}
        \State FAIL
      \Else
        \Comment{FAST meets the constraint}
        \If{$E(\textup{Oracle}) > T_E$}
          \Comment{Oracle does not meet the constraint}
          \State PASS
        \Else
          \Comment{Both FAST and Oracle meet the constraint}
          \If{$|E(\textup{Oracle}) - E(\textup{FAST})| < T_E$ \textup{and} $A(\textup{Oracle}, \textup{FAST}) < T_F$}
            \Comment{FAST is close enough to Oracle}
            \State PASS
          \Else
            \Comment{FAST is not close enough to Oracle}
            \State FAIL
          \EndIf
        \EndIf
      \EndIf
    \end{algorithmic}
    \caption{Verdict Expression Based on Oracle B}
    \label{fig:verdict-expression-b}
\end{algorithm}

\subsection{Test Suite}

Table~\ref{tab:test-suite} describes test cases that validate \FAST{}'s basic runtime and controller functionality. Test cases fall into two categories: those that validate normal system behavior (1-4), where the expected verdict is PASS, and those where the system executes outside of intended operating conditions (5-6), where the expected verdict is FAIL or INVALID.
In some tests, qualitative aspects of the application executions are interesting, beyond the verdict. For example, for test case 6, the adaptive execution's behavior should attempt to come as close to the goal as the available configurations allow.

\begin{table}[htbp]
    \centering
    \begin{tabular}{@{}lp{9.5cm}p{1.7cm}@{}}
        \toprule
        \# & Example & Expected \\
           & & Verdict \\
        \midrule
        1 & Change from low to high constraint goal (and vice versa) with enough time to meet the intent. & PASS \\
        2 & Change between constraint measures, e.g. from \code{latency} to \code{performance} (defined as 1/\code{latency}), while minimizing \code{energy}. & PASS \\
        3 & Change the optimization type from \code{min} to \code{max} while negating the objective function during the course of an execution. Should have no effect on the behavior of the system. & PASS \\
        4 & Negate the objective function should result in a change in the corresponding measure values. & PASS \\
        5 & Change from low to high constraint goal (and vice versa) without enough time to meet the intent. & FAIL \\
        6 & Use a constraint goal that is not achievable (no configuration achieves a constraint measure that is high/low enough). & INVALID \\
        \bottomrule
    \end{tabular}
    \caption{Application-Agnostic Test Suite for FAST}
    \label{tab:test-suite}
\end{table}

\section{Experimental Evaluation}
\label{sec:experimental-results}

\begin{table*}
    \begin{tabular}{
        @{}
        l
        p{4.1cm}
        p{2.0cm}
        p{2.1cm}
        p{0.9cm}p{1.1cm}p{0.7cm}
        @{\hspace{-0mm}}
    }
    \toprule
    \multirow{2}{*}{App.}
        & \multirow{2}{*}{Knobs}
        & Constraint
        & Opt. Type and
        & \multicolumn{3}{c}{Source Lines of Code}\\
        & 
        & Measure
        & Obj. Function
        & Base
        & Wrapper
        & FAST \\
    \midrule
    X264 
        & Motion Estimation Range, Sub-Pixel Refinement, Number of Reference Frames, Quantization Step 
        & \code{performance} 
        & max \code{quality}
        & 105992 & 296 & 58 \\
    SAR 
        & Coarse Decimation Ratio, Fine Decimation Ratio, Number of Ranges, Number of Beams 
        & \code{performance}
        & max \code{quality} 
        & 543 & -- & 41 \\
    Jacobi
        & Iterations Between Synchronization
        & \code{performance}
        & min \code{power} 
        & 165 & -- & 6 \\
    \bottomrule
    \end{tabular}
    \caption{Overview of applications used in experimental evaluation.}
    \label{tab:applications}

\end{table*}

\begin{figure*}[htbp]
    \centering
    \tiny
    \includegraphics{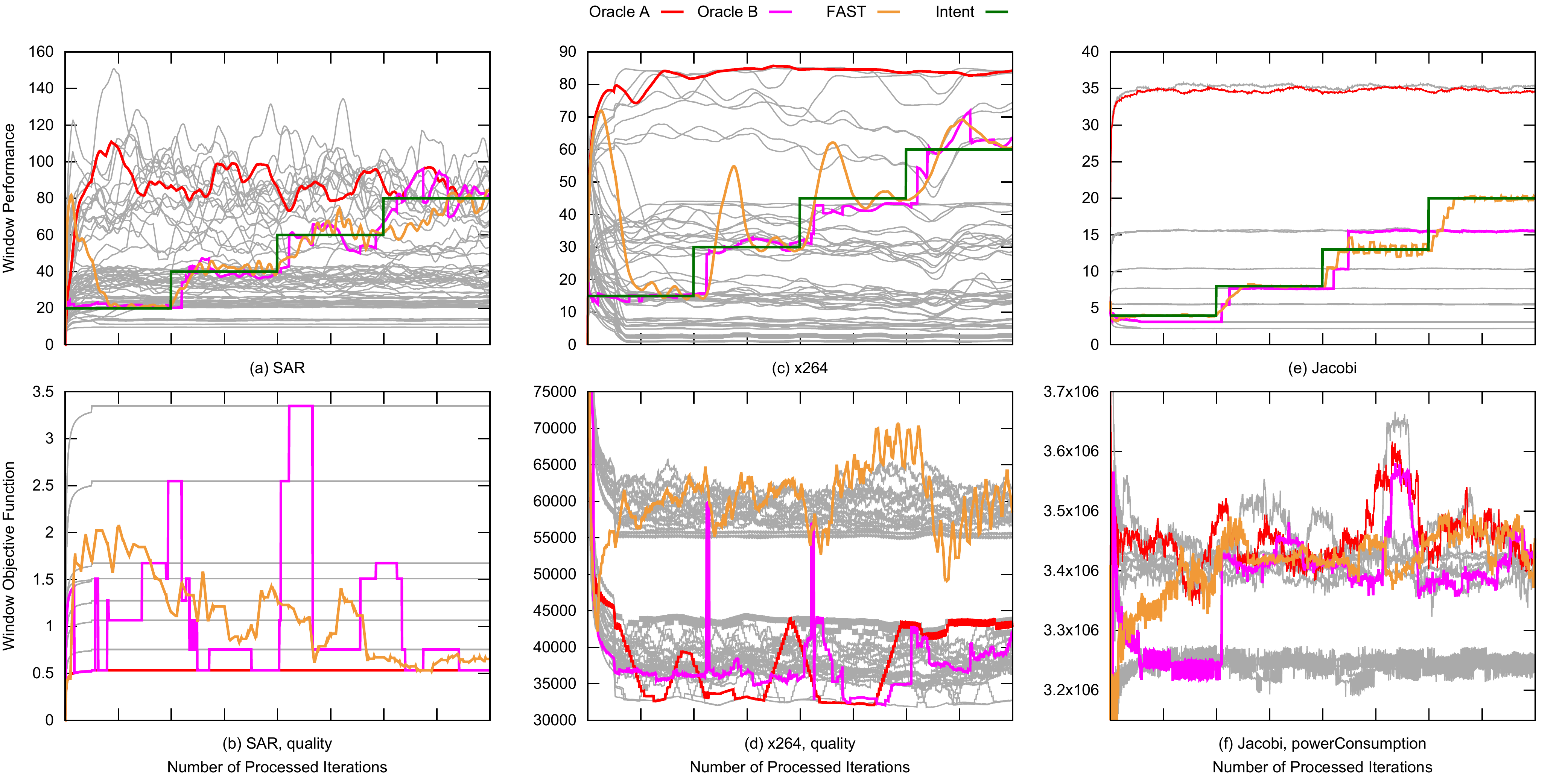}
    \caption{
        Comparison between FAST and oracles, showing adherence to intent, which is \code{max(quality) such that performance == g} for x264 and SAR, and \code{min(powerConsumption) such that performance == g} for Jacobi, where the constraint goal \code{g} is repeatedly perturbed during execution.
    }
    \label{fig:comparison-fast-oracles}
    \normalsize
\end{figure*}

Table~\ref{tab:applications} shows the three applications we use to evaluate \FAST{} including their knobs and  measures. The table also shows source--lines--of--code counts for each original (non-adaptive) application, wrapper code (when applicable), and \FAST{} instrumentation added to enable adaptation. As illustrated by these counts, software of any size can be controlled by \FAST{}.

Table~\ref{tab:knob-kinds-and-tradeoffs} further details each application's knobs. Knobs may be  \emph{ordinal}---such as the decimation ratio of the SAR application (Section~\ref{sec:radar})---or \emph{categorical}---such as the sub-pixel refinement method of the X264 application (Section~\ref{sec:x264}). All ordinal knobs discussed in this paper are discrete.

The cost of the optimization task that needs to be performed by the controller for every window varies across applications. For the constrained optimizing controller (Section~\ref{sec:constrained-optimizing-controller}), an optimal schedule will result in one change during the course of the window (of size $w$), and is a combination of two near-optimal configurations. The number of possible schedules is thus at most $N^2 * w$, where $N$ is the number of configurations in the controller model.

Figure~\ref{fig:comparison-fast-oracles} shows experimental results for the three example applications. The two plots for each application show window averages of the constraint measure (\code{performance}) and objective function (\code{quality}), to compare \FAST{} behavior to Oracles A and B. As summarized in Table~\ref{tab:comparison-fast-oracles}, these results illustrate that, by adapting dynamically to perturbations in the intent, \FAST{} consistently outperforms Oracle A's \code{quality}, while meeting the \code{performance} constraint (on average once the performance has stabilized after a perturbation). \FAST{} also performs on par with Oracle B and, interestingly, even beats it in terms of \code{quality}. This is possible because Oracle B is only an approximation of the ideal adaptive behavior, which meets the intent iteration-wise, choosing a single configuration at a time. \FAST{}, in contrast, interpolates between configuration over a window of inputs, allowing it to meet the \code{performance} constraint while staying in higher-\code{quality} configurations on average.

The fixed configurations (gray lines in Figure \ref{fig:comparison-fast-oracles}) show that the measures used to control the system (\code{performance} and \code{quality} in this case) can be highly noisy.
They fluctuate to the point where they may change order over the course of an execution. This change can cause a controller to oscillate, as the information provided by the controller's model conflicts with the feedback it receives. This situation is to be expected when intensional measures such as \code{performance} or \code{energy} are used for control, but it can also arise for extensional measures that are input-dependent, such as the x264 measure \code{quality} (as in Figure~\ref{fig:comparison-fast-oracles}c). Such behavior can be mitigated statically to some extent, by ensuring that the model contains configurations whose measures are sufficiently separated along the measures that will be used for control.

\begin{table}
    \centering
    \begin{tabular}{@{}l lll lll lll@{}}
        \toprule
        Statistic & 
            \multicolumn{3}{c}{x264} & 
            \multicolumn{3}{c}{SAR} & 
            \multicolumn{3}{c}{Jacobi} \\ 
         & 
            FAST & Or. A & Or. B &
            FAST & Or. A & Or. B &
            FAST & Or. A & Or. B  
            \\
        \midrule
        %
        %
        %
        %
        MAPE &
            0.032 & 0.941 & 0.025 & 
            0.027 & 0.667 & 0.036 & 
            0.008 & 1.678 & 0.085 \\ 
        Mean Obj. Function & 
            6.18e4 & 3.80e4 & 3.60e4 &  
            1.10   & 0.53	& 0.96   &  
            3.42e6 & 3.46e6 & 3.39e6 \\ 
        \bottomrule
    \end{tabular}
    \caption{Summary of Application Executions Comparing FAST to Oracles A and B}
    \label{tab:comparison-fast-oracles}
\end{table}

\begin{table}
    \centering
    \begin{tabular}{@{}llll@{}}
        \toprule
        Module & Knob & Kind & Key Measure Trade-Off \\
        \midrule
        
        System & Utilized Cores & Ordinal &
            \code{latency} $\updownarrows$ \code{energy} \\
        & Utilized Core Frequency & Ordinal &
            \code{latency} $\updownarrows$ \code{energy} \vspace{1mm}\\
            
        X264 & Motion Estimation Range & Ordinal &
            \code{latency} $\updownarrows$ \code{bitrate} \\
        & Sub--Pixel Refinement & Categorical &
            \code{latency} $\updownarrows$ \code{bitrate} \\
        & Number of Reference Frames & Ordinal &
            \code{latency} $\updownarrows$ \code{bitrate} \\
        & Quantization Step	& Ordinal & 
            \code{latency} $\updownarrows$ \code{quality} \vspace{1mm}\\
            
        SAR & Coarse Decimation Ratio & Ordinal &
            \code{latency} $\updownarrows$ \code{quality} \\
        & Fine Decimation Ratio & Ordinal &
            \code{latency} $\updownarrows$ \code{quality} \\
        & Number of Ranges & Ordinal &
            \code{latency} $\updownarrows$ \code{quality} \\
        & Number of Beams & Ordinal &
            \code{latency} $\updownarrows$ \code{quality} \vspace{1mm}\\
            
        Jacobi & Iterations Between Synch. & Ordinal &
            \code{latency} $\updownarrows$ \code{powerConsumption} \\
            
        \bottomrule
    \end{tabular}
    \caption{Application and Platform Configuration Knob Kind and Trade-offs}
    \label{tab:knob-kinds-and-tradeoffs}
\end{table}

\subsection{Synthetic Aperture Radar (SAR)}
\label{sec:radar}

SAR is a signal processing pipeline that detects objects in a sequence of synthesized signals. Both the size of filters (Coarse Decimation Ratio, Fine Decimation Ratio) and granularity (Number of Beams, Number of Ranges) are tunable knobs. While SAR and the numerical application described below were implemented in pure Swift, SAR is the only one that relies on Grand Central Dispatch to run sub-tasks in parallel due to its pipelined nature.

\subsection{Jacobi Iterative Method}

The Jacobi Iterative Method is an algorithm that approximates the solution of a diagonally dominant system of linear equations. We implemented a parallel version of the algorithm in Swift. The parallelism was constructed using a POSIX style threading system supported in Swift. During execution, in addition to the number of available cores and the core frequency, the controller can choose the number of iterations after which the threads should synchronize. The application is non-trivial to synchronize, because reducing the number of iterations after which to synchronize changes the overall time to reach convergence and the speed of meaningful iterations. This application illustrates the generality of \FAST{}, in that it works equally well to control streaming and computation bound applications along with applications that have different resource utilization patterns.  

Furthermore, the Jacobi Iterative Method experiments along with the x264 experiments show several advantages over auto-tuning systems such as OpenTuner. While dynamic adaptation is the most obvious advantage, since OpenTuner provides a single configuration in which to run the application throughout the entirety of the execution, \FAST{} allows knobs to be dynamically set in response to environmental fluctuation. Since OpenTuner provides a single configuration in which to run the application, there might not exist a configuration using which the application can achieve the required intent, this outlines another advantage of \FAST{} over OpenTuner, since \FAST{} can still achieve the intent by switching between an under-provisioned and an over-provisioned configuration in an execution window. This capability is illustrated in Figure~\ref{fig:comparison-fast-oracles}(e), where the two last performance goals (13 and 20) are not close to any fixed configuration's performance, but can be met by \FAST{} on average by interpolating between configurations. Similarly, in application such as x264, in which performance is highly dependent on the input, there can be a significant amount of variation in performance using a single configuration, \FAST{} mitigates this using feedback control. 

\subsection{x264 Video Encoder}
\label{sec:x264}

Modern video encoding---represented by x264~\cite{merritt2006x264}---is a quintessential streaming application, exhibiting several characteristics that make it a useful test case for an intent-driven programming. It exposes a myriad of parameters and measures with different trade-offs, making up-front configuration of this application something of an art~\cite{x264-settings,x264-subme-comparison}. For the purpose of evaluation, the x264 C++ code base was instrumented using the Swift foreign-function interface. The x264 application uses a typical high--level pipeline for signal processing. We have exposed four of its application parameters as knobs for testing purposes. 

In the following subsection we present experimental results for an example use case based on the FAST library API functions \code{restrict} and \code{control}.
The example was implemented as an extension of the x264 application, and shows that the system can be adapted to suit the needs of a diverse range of real-world
application using minimal developer effort. 

\subsubsection{Use Case: Explicit Higher Quality Demand}
\label{sec:x264-use-case-explicit-higher-quality-demand}

Developers might explicitly want a range of frames to be encoded at a significantly higher quality, while still maintaining the original intent. The \code{Knob} type's \code{restrict} method enables this by setting a knob that affects quality to a particular value or range (e.g., of higher motion estimation or constant quantizer). As an example, consider footage produced by a CCTV camera. For most of the day, nothing of interest is captured and, hence, low quality footage suffices. During times of interest, the developer might disallow configurations that produce low quality video. To this end, the developer can use the \code{restrict} API call to set the
range of values that a knob can be assigned to get higher quality frames. With these restrictions in place, 
the controller uses the remaining knobs to meet the intent. Once the
time of interest has passed, the knob restrictions can be lifted (using the \code{control} method) and the controller is free to use the
complete range of values listed in the intent specification.

\begin{figure}
    \centering
    \Small
    \hspace*{-6.5mm}\includegraphics[width=0.75\textwidth]{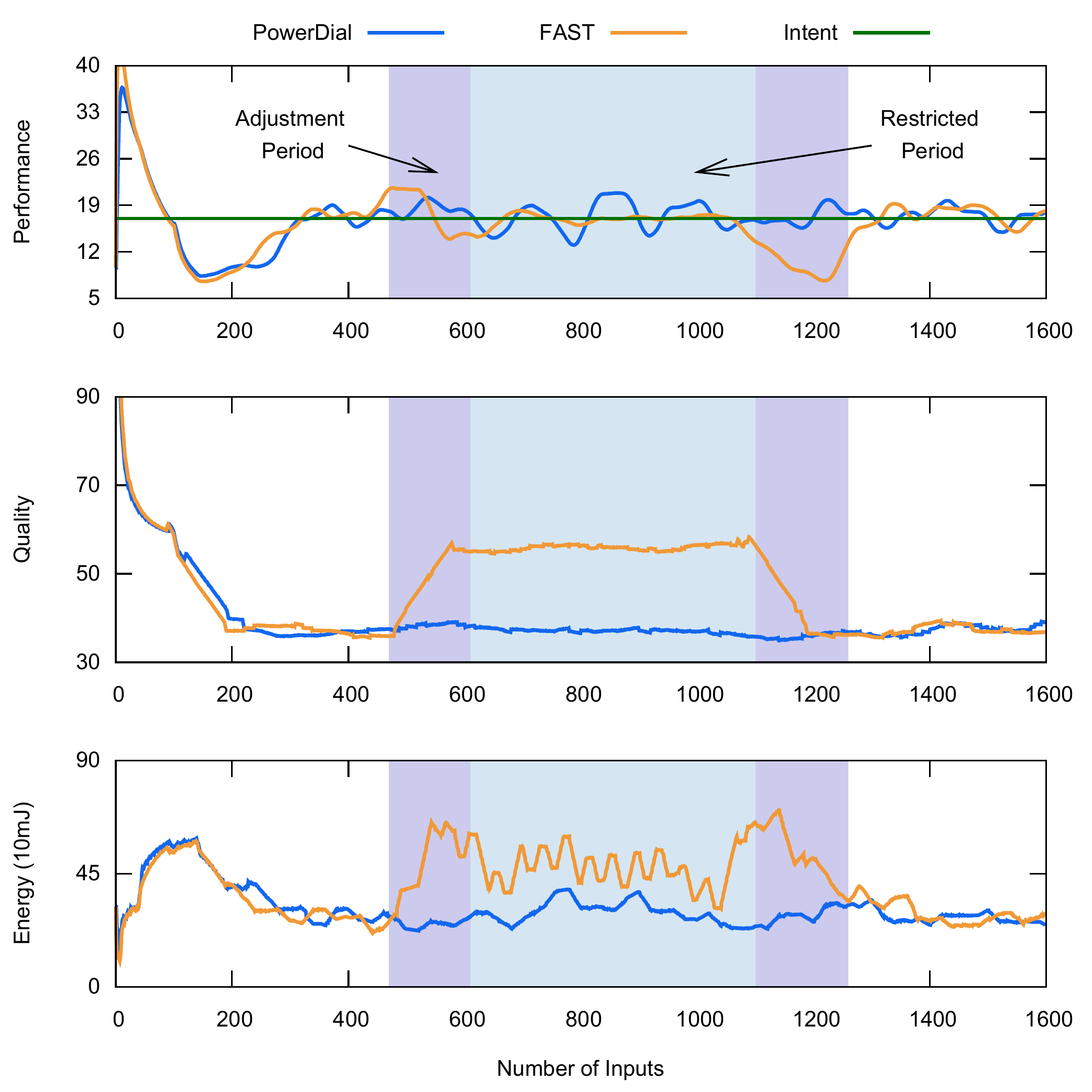}
    \caption{CCTV example execution measures.}
    \label{fig:cctv-example}
\end{figure}

Figure~\ref{fig:cctv-example} shows the execution of such an application with the following intent: 
\[
\code{min(energyPerFrame) such that performance == 17.0}
\]
The controller minimizes energy per frame until the execution reaches the range of inputs that is of special interest to the developer. Via a call to the \code{restrict} method the developer sets a restriction on the range of values that a particular \code{Knob} can take. Consequently, the controller is forced to choose from among the configurations that are chosen by the developer (which produce higher quality frames). 
Figure~\ref{fig:cctv-example} shows that the frames in the restricted range are encoded at a higher quality than the rest of the frames, while still meeting the constraint, but that a higher amount of energy is required to encode each frame over this period.

While the execution is in this restricted range the controller intelligently copes with the change by selecting values for the other knobs so as to be able to meet the constraint. For example, the controller chooses configurations with a greater number of cores and higher core frequencies which results in an increase in energy required to encode a frame as shown in Figure~\ref{fig:cctv-example}. 

During this time the controller still meets the intent while providing higher quality at the cost of a higher energy to encode a single frame. Once the inputs in the range of interest have been processed and the restriction on the \code{Knob} values has been lifted using a call to the \code{control} method, the controller can return to choosing \code{Knob} values that minimize energy while meeting the performance constraint (i.e. minimize objective function under the active constraint). 

After the call to \code{control} the state of the controller is reset. Hence, the controller takes a small period to adjust to the new configurations. 
Implementation of an application with such requirements requires no more than two API calls to the \FAST{} runtime. 

This example illustrates a capability that is easy to achieve in \FAST{}, but difficult using prior work.  The PowerDial System adjusts knob configurations dynamically to meet a target performance despite environmental fluctuations \cite{hoffmann2011dynamic}, but it provides no mechanism for the program to change the available set of knobs.  In fact, changing available knobs in PowerDial would require recompilation, while in \FAST{} it requires only a few method calls. Thus, this example demonstrates an advantage of the combination of procedural and continuous control of adaptation that is possible in intent-driven programming. 
The small amount of changes needed to incorporate this complex capability in an application using \FAST{} also demonstrates a key usability advantage over prior work.


\section{Conclusions and Future Work}
\label{sec:future-work}

This paper presents the \FAST{} architecture for intent-driven programming, and describes some fundamental aspects of computing that are affected by extending a language with intents. This extension reopens problems for which reasonable solutions are available for traditional languages, such as what kinds of programming language constructs are useful, how to formalize the semantics of the language, and how to validate program correctness. Answering these questions is part of our future work. 

Specifically, we plan to extend the language with a richer, composable syntax and more general optimization to support multi-objective and multi-constraint intents, as well as additional knob kinds, such as dense ordinals. Such extensions will require a controller component with the corresponding capabilities. For example, support for multi-constraint intents could be added using algorithms for solving mathematical programming problems, such as the simplex or interior-point methods. Notably, however, such a change would not require any changes to the programming model. Generalizing the knob and measure concepts, for example to accommodate input features, will improve the ability of \FAST{} to control applications whose inputs exhibit distinct phases. Runtime extensions to support active profiling will also be helpful in that regard.

We also plan to explore the topic of correctness of intent-driven programs, including static and dynamic analyses that check that assumptions under which components operate correctly (such as various properties of the controller model) are not violated. Leveraging statistical information in the runtime will make it possible to monitor and predict the violation of more general assurance criteria.


\bibliography{main}

\end{document}